\documentclass[twocolumn,eqsecnum,aps]{revtex4} 
\usepackage{amsmath}
\usepackage{latexsym}
\begin{document}
\title{ The Casimir Effect on the Light-Cone}
\author{Frieder Lenz and Daniela Steinbacher}
\address{Institut f\"ur Theoretische Physik III\\ 
Universit\"at Erlangen-N\"urnberg\\ 
Staudtstra\ss e 7\\ 
D-91058 Erlangen\\ 
Germany }
\date{\today} 
\begin{abstract}
The Casimir effect is investigated in light-cone quantization. It is shown that for spacelike separation of the walls enclosing the system the standard result for the pressure exerted on the walls is obtained. For walls separated in light-cone space direction no regularization of the  quantum fluctuations exists which would yield a finite pressure. The origin of this failure and its implications for other vacuum properties are discussed  by analyzing the Casimir effect as seen from a moving observer approaching the speed of light. The possibility  for calculation of thermodynamic quantities in light-cone quantization via the Casimir effect is pointed out.   
\end{abstract} 
\maketitle


\section{Introduction}
In the Casimir effect \cite{casimir}, the change in the quantum fluctuations of a  field due to its interaction with a medium incorporated into boundary conditions is measured. The observable is the  pressure exerted by the quantum fluctuations on  walls which limit the system in one spatial direction. The Casimir effect is accessible to experiments only if the corresponding quantum field possesses massless excitations. Measurements \cite{sparnaay,roy,ono} of the change in the ground state energy of the electromagnetic field in the presence of metallic boundaries have confirmed Casimir's original prediction. In the standard treatment of the Casimir effect the appropriate standing wave conditions of  electromagnetism are used. In relativistically covariant theories, the Casimir effect with periodic boundary conditions imposed and black-body radiation are related to each other. More precisely, covariance connects the energy-momentum tensor for a system at finite spatial extension with the energy-momentum tensor of the same system at finite temperature \cite{lt98,toms}. \\
This investigation of the Casimir effect  is intended to clarify the description of the vacuum in light-cone quantization. Unlike other vacuum properties like condensates which have to  appear for consistency of the underlying theory  but cannot be measured directly, quantities related to the Casimir effect are experimental observables and therefore have to be correctly described within any formalism.
As in other instances, one might expect that the infinite momentum frame interpretation of light-cone results applies. In this case the light-cone formulation of the Casimir effect should correspond to the observation of the Casimir effect by an observer moving with respect to the ``walls'' of the system and approaching the speed of light. Of particular interest thereby is the situation in which the observer's velocity is perpendicular to the ``walls''.
 With this study of the light-cone vacuum properties we will also address the issue of developing a viable approach to finite temperature field theory in  light-cone quantization.\\
Our investigation will start within the canonical formalism. In this standard framework for the discussion of the Casimir effect \cite{itzykson} energy density and related observables are obtained by the (suitably regularized) sum over the zero-point energies of the normal modes of a massless non-interacting quantum field. As one of the central issues in our study we will  establish the relation between the Casimir energy in ordinary coordinates and on the light-cone  within a covariant formalism and prepare in this way the ground for the discussion of the relation to finite temperature field theory on the light-cone.


\section{Casimir Effect in the Canonical Formalism}  
The forces acting on the boundaries of an (partially) enclosed system are determined by 
the size dependence of the  energy density. In this section we shall calculate this energy 
density for periodic boundary conditions. Although not directly relevant for the observation 
of the Casimir effect in electrodynamics, where standing wave boundary conditions 
describe appropriately the interaction of electromagnetic waves with metallic boundaries, 
from the theoretical point of view, periodic or antiperio\-dic boundary conditions are preferable. 
Momentum conservation is preserved with this choice and for re\-la\-tivistically covariant theories, 
the results  can be connected to the corresponding thermodynamic quantities at 
finite temperature. Here we discuss the Casimir effect for a non-interacting, massless scalar field. In this section,  heat-kernel regularization is used for dealing with the infinities in the sum over the zero-point energies. For comparison 
we first give the result for the energy density using standard coordinates. Enclosing  the system 
bet\-ween walls at a distance $L$ and imposing periodic boun\-dary conditions the 
eigenenergies of the one particle states are   
$$\omega({\bf k}_{\perp},n) = \sqrt{{\bf k}_{\perp}^2 +\left(\frac{2\pi n}{L}\right)^2}$$
with ${\bf k}_{\perp}$ denoting the continuous momentum components orthogonal to the compact 
direction. As is well known, the regularized sum over the zero-point energies  
\begin{eqnarray}
\label{energie}
\langle{\cal H}_{\lambda^{0}}\rangle &=& \frac{1}{2L} \int\frac{d^2{\bf k}_\perp}{(2\pi)^2}\sum_{n=-\infty}^{\infty}\omega \,e^{-\lambda^{0}\omega}\nonumber\\
&=& \frac{3}{2\pi^2\lambda^{0\,4}}-\frac{\pi^2}{90 L^4} 
\end{eqnarray}
contains a singular, size independent contribution and a finite, size dependent term. 
The observable pressure exerted on the walls 
$$P= -\frac{\partial \langle L {\cal H}_{\lambda^{0}}\rangle }{\partial L} =- \frac{\pi^2}{30 L^4} $$
is therefore finite and  its $L$-dependence follows from dimensional arguments.
\newline
We now calculate the energy density in light-cone quantization. We use the following notation for 
coordinates and momenta
$$x^{\pm}=\frac{1}{\sqrt{2}}(x^{0}\pm x^{3}),\quad k_{\pm}=\frac{1}{\sqrt{2}}(k_{0}\pm k_{3})$$
and refer to $x^{+},k_{+}$ as light-cone time and light-cone ener\-gy respectively. 
The light-cone energies of the one par\-ticle states are given by
$$k_{+} = \frac{k_{1}^2+k_{2}^2}{2 k_{-}}\, , \quad \mbox{with the constraint} \quad k_{-} > 0 \, .$$ 
In light-cone quantization  the system may be chosen to be compact and periodic  
in a transverse direction (orthogonal to the 3 direction) or in the light-cone space $x^{-}$ direction. 
With the transverse boundary condition
$$\varphi(x^+,x^-,x^1+L,x^2)= \varphi(x^+,x^-,x^1,x^2) \, ,$$
the energies of the one particle states are
$$\omega_{t}(k_-,k,n)=\frac{1}{2 k_{-}}\left(k^2+\left(\frac{2\pi n}{L}\right)^2\right).$$
To regularize the infinities the suppression of the contribution from both large light-cone energies and large  
light-cone momenta requires  two regulators. The resul\-ting 
light-cone energy density  
\begin{eqnarray}
\label{lct}
\langle{\cal H}_{\mbox{t}}\rangle&=&\frac{1}{2L}
\int_{0}^{\infty}\frac{ dk_-}{2\pi}\int_{-\infty}^{\infty}\frac{dk}{2\pi}\sum_{n=-\infty}^{\infty} \omega_{t}\,e^{-\lambda^{-}k_--\lambda^{+} \omega_{t}}
 \nonumber\\ &=&\frac{1}{8\pi^2 (\lambda^{+}\lambda^{-})^2} - \frac{\pi^2}{90 L^4}
\end{eqnarray}
coincides in the relevant, finite and size dependent term and differs in the singular but size 
independent contribution. Later we will make explicit the relation between the results. 
  \newline
With longitudinal boundary conditions
$$\varphi(x^+,x^-+L,x^1,x^2)= \varphi(x^+,x^-,x^1,x^2)$$
the energies of the one particle states are
$$\omega_{l}({\bf k}_{\perp},n)=\frac{{\bf k}_{\perp}^2}{4 \pi n/L} \,,$$
and the following result for the energy density 
\begin{eqnarray}
\label{lcl}
\langle{\cal H}_{\mbox{l}}\rangle&=&
\frac{1}{2L}\int\frac{d^{2}{\bf k}_\perp}{(2\pi)^2}\sum_{n=0}^{\infty}\omega_{l}
e^{-\lambda^{-}2\pi n/L-\lambda^{+}\omega_{l}} \nonumber \\
&=& \frac{1}{8\pi^2 (\lambda^{+}\lambda^{-})^2} - \frac{1}{24\lambda^{+\,2} L^2} + 
\frac{\pi^2}{120 L^4}\left(\frac{ \lambda^{-}}{\lambda^{+}}\right)^{2}\nonumber \\
\end{eqnarray}
is obtained. No separation of regulator ($\lambda^{\pm}$) and size  ($L$)  dependence occurs. 
 It is difficult to assess the physical relevance of this result within the canonical formalism.  In the following sections we will recalculate the 
Casimir energy density in 
a formalism in which the (residual) covariance is explicit.


\section{Energy Momentum Tensor in a Periodic Vacuum}
To make explicit  the covariance in the calculation of the Casimir energy we impose the boundary condition  
\begin{equation}
  \label{bc0}
  \mbox{bc}:\quad \varphi (x+\ell) = \varphi (x) \, .
\end{equation}
The 4-vector $\ell$ specifies orientation and distance of the ''walls'' enclosing the system. Imposing this boundary condition singles out a Lorentz-frame which we will refer to as the rest-system.  
The generating  functional in a frame 
connected with the rest-system by an element $\Lambda$ of the (proper) Lorentz-group is given by
$$
Z _{\Lambda}\, [J] = \int_{\mbox{bc}_{\Lambda}} \, d \, [\tilde{\varphi}]  \; e^{i \int \, d^{4}\, x \; {\cal L} (\tilde{\varphi} (x)) + i \; \int \, d^{4}\, x J \cdot \tilde{\varphi}}
$$
with the  Lorentz-transformed boundary conditions  bc$_{\Lambda}$ 
\begin{equation}
  \label{bc1}
\mbox{bc}_{\Lambda}:\quad \tilde{\varphi} (\Lambda ^{-1}(x + \ell) ) = 
 \tilde{\varphi} (\Lambda^{-1} (x )) \,.
\end{equation}
We perform a variable substitution  corresponding to a 
Lorentz-transformation   back to the rest-system
$$ \; \varphi(x) = \tilde{\varphi} (\Lambda^{-1} x)$$
The Lagrangian  ${\cal L}$ is invariant, the Jacobian of this variable substitution is $1$ and the 
fields $\varphi$ satisfy the boun\-dary conditions (\ref{bc0}).
The generating functional for a  moving observer can therefore be written as 
$$
Z _{\Lambda}\, [J] = \int_{\mbox{bc}} \; d \; [ \varphi ] \; e^{i \int \, d^{4}\, x \; 
{\cal L} (\varphi(x)) + i \int d^{4} x \; J (\Lambda^{-1} \,x) \; \varphi (x)} \, .$$
With the Lagrangian of a massless non-interacting scalar field
$$ \; {\cal L}[\varphi] =  \frac{1}{2}\,\partial _{\mu} \varphi \, \partial ^{\mu} \varphi \,,$$ 
the generating functional becomes
$$
Z_{\Lambda}\, [J] =e^{- \frac{i}{2} \, \int \, d^{4} x \, d^{4} y \, J(x) \, D(\Lambda (x-y)) \, J(y)}$$
with the scalar two-point function periodic in the direction of $\ell$
\begin{equation}
  \label{d3}
D (z) = \sum_{n} \, \int \, \frac{d^{4}k}{(2 \pi )^{3}} \,e^{i k z}
\; \frac{1}{k^{2} + i \epsilon}\, \delta \left(k\ell - \,2 \pi n
\right ).  
\end{equation}
For evaluation of the Casimir effect as seen from a moving observer we compute 
the energy-momentum tensor with the help of the generating functional 
\begin{eqnarray}
 \label{emt}
\langle J_{\mu\nu}\rangle &=& -(\partial_{\mu}^{x}\partial_{\nu}^{y}-\frac{1}{2}
g_{\mu\nu}\partial_{\rho}^{x}\partial^{y\,\rho})\frac{\delta ^{2}Z_{\Lambda}}
{\delta J(x)\delta J (y)}\Big|_{J = 0,\; x \rightarrow y}\nonumber\\ &=& 
-i(\partial_{\mu}\partial_{\nu}-\frac{1}{2}g_{\mu\nu} \Box)
D(\Lambda z)\Big|_{z \rightarrow 0} \, . \, \nonumber\\ 
\end{eqnarray}
We note that in this framework the ultraviolet divergencies appearing in the Casimir energy are regularized by point splitting.  The covariance of this regularization  will turn out to be  crucial for the following  studies.

 In the evaluation of  $D(z)$ we have to distinguish the cases of space- and lightlike 4-vectors $\ell$.  We first consider spacelike separations of the walls 
 \begin{equation}
  \label{subst}
0<  -\ell^2 = L^{2} ,  
\end{equation}
and define correspondingly the components of the 4-vector $z$ which serves as regulator
\begin{equation}
  \label{coz}
 z^{\|}= \frac{ z\ell }{L} \,\quad z^{\perp} = \sqrt{z^{2}+z^{\|\, 2}-i\epsilon} \;.  
\end{equation} 
  Integration over the momenta and summing the resulting geometrical series  yields the final expression for the two-point function  
\begin{eqnarray}
\label{DL}
  D (z) =  - \frac{1}{4\pi L z^{\perp}}\Big [ 1 &+&
\frac{1}{ e^{2i \pi /L (z^{\perp} + z^{\|}) }-1} \nonumber\\ &+& 
\frac{1}{ e^{2i \pi /L (z^{\perp} -  z^{\|}) }-1}\Big]\; .
\end{eqnarray}
The above expression displays the covariance of this approach. In general,  the scalar two-point function is a function of the scalars formed from the two 4-vectors $z$ and $\ell $ characterizing the system
\begin{equation}
  \label{invd}
  D(z)= D(z^2,\ell ^2, z\ell\, )\;. 
\end{equation}


\section{Casimir Effect in a Moving Frame}
In this section we compute   the components of the energy-momentum tensor
in various systems.  We expand the 
above expression for $D(z)$ (Eq.(\ref{DL})) around $z=0$ 
\begin{eqnarray}
  \label{expa}
  D(z)& = &  -\frac{1}{4\pi L z^{\perp}}\Bigg\{ 1 + \sum^{\infty}_{n=0} \, \frac{B_{n}}{n !}
\left( \frac{2i\pi}{L}\right)^{n-1}\nonumber\\& \cdot& \bigg [\left ( z^{\perp} +  z^{\|} \right )^{n-1} +  \left (z^{\perp} -  z^{\|} \right ) ^{n-1} \bigg ]\Bigg\}
\end{eqnarray}
with the Bernoulli-numbers $B_{n}$. The leading terms in the $ z \rightarrow 0$ limit
are
$$
  D (z)\approx \frac{i}{4\pi} \; \left [ \, \frac{1}{\pi z^2} - \frac{\pi}{3 L^{2}}\, -
\, \frac{\pi^{3}}{45 L^{4}} \; \left ( z^{2} +4 \,z^{\|\, 2}\right ) \right ].
$$
The singular term in this expansion is invariant under Lorentz-transformations 
($\Lambda$) and independent of the periodic structure of the vacuum ($L$). At the smallest scale, 
the vacuum is identical for all observers and not affected by periodicity on large scales. The  
singular contributions to the energy-momentum tensor is given by 
\begin{eqnarray}
\label{sing}
 \langle J_{\mu\nu}^{sing}\rangle &=& 
(\partial_{\mu}\partial_{\nu}-\frac{1}{2}g_{\mu\nu}\partial_{\rho}\partial^{\rho})\frac{1}{4\pi^2 z^2} \nonumber\\ &=& 
\frac{1}{2\pi^2}\left[\frac{4z_{\mu}z_{\nu}}{z^6}-g_{\mu\nu}(\frac{1}{z^4}+i\pi^2\delta(z))\right].
\end{eqnarray}
By choosing the elements of the energy-momentum tensor to vanish in a certain frame 
and for a certain size, e.g. the rest-system and  $L=\infty$ , the singular pieces will be absent 
in any other frame and for any other $L$. It does not affect any observable. For the regular, size and frame dependent part of the energy-momentum tensor we obtain   
\begin{equation}
  \label{energy2}
\langle J_{\mu\nu}\rangle = - \,\frac{\pi^2}{90 L^4}\, \Lambda_{\mu}^{\rho} \,
\Lambda_{\nu}^{\sigma} \,\left[g_{\rho\sigma}-4 \,\frac{\ell _{\rho}\, \ell _{\sigma}}{\ell^2}\right]\; .\end{equation}
The energy density in the rest-system  coincides, after  the identification
 $z_{\mu}=i\lambda_{0}\delta_{0\mu}$, with the result of the canonical calculation (Eq.(\ref{energie})) 
 up to the $\delta(z)$ term which disappears in the rotation to the Euclidean space. 

The light-cone energy density in the transverse Casimir effect ($\Lambda =1\, , \l^{\mu}= \delta_{\mu,1}$)  is given by 
\begin{equation}
  \label{lced}
  \langle J_{+ - }\rangle = \frac{1}{2}\langle J^{\,00} - J^{\,33} \rangle = - \frac{\pi ^{2}}{90 \, L^{4}} \, .
\end{equation}
 With the identification
 $z^{\pm}=i\lambda^{\pm}, z^1=z^2=0$ it agrees   up to the $\delta$-function with the 
canonical result (Eq.(\ref{lct})). 
In this covariant formulation,  the evaluation of $D(z)$ is trivially identical for light-cone and ordinary 
coordinates. The difference in the singular piece arises from the differences in the definition 
of $J_{+-} $ and $J_{00}$.  \\
In the longitudinal Casimir effect the choice of light-cone coordinates is much more severe. Here, the 4-vector $\ell$ is light-like. A simple relation between the  two-point functions  for  spacelike and lightlike separations of the compact direction does not exist. 
One however might expect  
the  light-cone energy density in the longitudinal Casimir effect  to be related 
with  the energy density as seen from an observer in the infinite momentum frame. We will investigate this possibility and assume in the following discussion  the 3-direction  to be compact
$$\ell ^{\mu} = L\delta_{\mu 3}$$ 
and  $\Lambda$ to describe a boost in the 3-direction 
\begin{equation}
  \label{boo}
\Lambda^{3}_{\mu} = \gamma (\delta_{\mu 3}-\beta \delta_{\mu 0})\;.  
\end{equation}
 
 The order in which the  infinite momentum frame  limit $\quad \beta^2 \rightarrow 1$ and  the limit of vanishing regulator $z$ are performed has to be specified. If we first perform for given $\beta^2 < 1$ the $z \rightarrow 0$ limit $ |(\Lambda z)^{\mu}| \rightarrow 0$  we can expand as above  the exponentials in Eq.(\ref{DL}) and   the result (cf. Eq.(\ref{energy2})) is given by  
 \begin{equation}
  \label{edmf}
  \langle {\cal H}\rangle  = -\frac{\pi^{2}}{90 \, L^{4}} \; \left [1 + 4 \beta^{2} \gamma ^{2} \right ] , 
\quad \langle J_{33}\rangle =  - \frac{\pi^2}{90 L^{4}} \gamma ^{2}(3 + \beta ^{2} ).
\end{equation}
In the subsequent $\beta^2 \rightarrow 1$ limit, the elements of the energy-momentum tensor become infinite, the  light cone energy density however 
\begin{equation}
  \label{lced2}
  \langle J_{+ - }\rangle = \frac{1}{2} \left (J^{\,00} - J^{\,33} \right ) = \frac{\pi ^{2}}{90 \, L^{4}}\, ,
\end{equation}
by covariance,  is independent of $\beta$ and thus is not affected by the approach to the infinite momentum frame. This result does 
not agree  with the canonical result (Eq.(\ref{lcl})). It differs in sign from the light-cone energy density in the transverse Casimir effect due to the extra momentum flux (pressure) along the compact direction induced by the walls contributing here but not in (\ref{lced}). \\ 
In the reversed order, first the  approach to the light-cone is performed (for fixed regulator $z$) and only then  the $z \rightarrow 0$ limit is carried out. 
The transformed arguments (cf. Eqs.(\ref{coz}), (\ref{boo})) appearing in the expression (\ref{DL}) for the two-point function are to leading order in the  $\beta \rightarrow -1$ limit  given by 
\begin{eqnarray*}
\Lambda (z^{\perp}+ z^{3}) &\approx& \left[\,\frac{2}{\sqrt{1+\beta}} \; z^{+} - \,
\frac{z^{2}_{\bot} \sqrt{1+\beta}}{2 z^{+}}\,\right]\nonumber\\
 \Lambda (z^{\perp}- z^{3}) &\approx& \frac{\sqrt{1+\beta}\,  z^2}{2z^{+}}\;.
\end{eqnarray*}
Keeping 
\begin{equation}
  \label{un1}
|z^2| \ll L^2 , \quad z^{\perp\, 2} \ll L^2
\end{equation}
fixed and approaching the light-cone ($\beta \rightarrow -1$),  the quantity 
\begin{equation}
  \label{Lz}
\tilde{L}^2 = \frac{ z^{+} L}{\sqrt{1+\beta}} 
\end{equation}
increases beyond any bound and becomes the largest characteristic length of the system  
\begin{equation}
  \label{un2} |z^2| \ll  L^2 \ll \tilde{L}^2 \;.
\end{equation}
In this limit the following expression for the 2-point function 
\begin{equation}
  \label{DL2}
  D (\Lambda z) = -\frac{1}{4 \pi \tilde{L}^{2} }+ D _{+} (\Lambda z) + D_{-} (\Lambda z) 
\end{equation}
is obtained, with 
\begin{equation}
  \label{DL2+}
  D_{+} (\Lambda z) \approx -\frac{1}{4 \pi \tilde{L}^{2} } \,
\frac{1}{ \exp\{i\pi\frac{4 \tilde{L}^{2} }{L^{2}}\}-1} \;,
\end{equation}
\begin{eqnarray}
\label{exp}
 &&D_{-} (\Lambda z) \approx -\frac{1}{4 \pi \tilde{L}^2} \frac{1}{ \exp\{i \pi \frac{z^{2}}{\tilde{L}^2}\}-1 } \nonumber \\
&&\approx \frac{i}{4 \pi ^{2} z^{2}} +\frac{1}{8 \pi \tilde{L}^{2} }    -i \frac{ z^{2}}{48\, \tilde{L}^{4}}-i\frac{ \pi ^{2} z^{6}}{2880 \, \tilde{L}^{8}} \;. 
\end{eqnarray}
Note that the argument of the exponential in $  D_{+} $ becomes infinite in the approach to the light-cone. After a rotation to complex $z$  the  contribution of  $  D_{+} $ to the energy-momentum tensor becomes negligible. The argument of the exponential in $  D_{-} $ remains small and the result for the  energy density in the infinite momentum frame follows  
\begin{eqnarray}
  \label{lcl2}
&&\langle {\cal H}_{lc} \rangle = \langle J_{+-} \rangle = - \; \frac{i}{2} \,\mbox{\boldmath$\nabla$}^{2}_{\perp}\,D (\Lambda z) \nonumber \\
&\approx& -\frac{1}{2\pi^2 z^4}
(1-\frac{4z^{+}z^{-}}{z^2})+ \frac{1}{24 \tilde{L}^{4}}+\frac{\pi^{2}z^{2}}{480 \tilde{L}^{8}}(3z^{2}-4z^{+}z^{-}) . \nonumber\\
\end{eqnarray}
In the infinite momentum frame limit specified by Eqs. (\ref{un1}), (\ref{un2}), the $\tilde{L}^{-8}$  and the higher order contributions in the above formulae can be neglected. Identifying the regulators in the canonical with those of the infinite momentum frame calculation
$$ z^{+}= i\lambda^{+}(1+\beta)^{\frac{1}{2}}\,\quad z^{-} = -i\lambda^{-}(1+\beta)^{-\frac{1}{2}},\quad {\bf z}_{\perp}=0$$ 
makes the  expressions in Eqs.(\ref{lcl}) and (\ref{lcl2}) for the longitudinal Casimir energy density coincide.  \\
In performing first the limit $ \beta^2 \rightarrow 1$   necessarily  the regime specified by  Eq.((\ref{un2})) is reached in which the Lorentz-contracted length becomes much smaller than the regulator 
$$ \gamma^{-1} L \ll |z^{+}| \;.$$ 
A regulator independent result for the physical observables cannot be expected in such a situation.


\section{Lightlike and Spacelike Compactifications}
 For lightlike orientation of the compact direction,  
$$\ell = \frac{L}{\sqrt{2}}(1,0,0,-1)$$
the  two point function (\ref{d3})  can be evaluated with the result
\begin{eqnarray}
  \label{dlc1}
D_{lc} (z) &=& -\,\frac{1}{4 \pi z^{+} L} \frac{1}{ e^{i \pi \frac{z^{2}}{z^{+} \, L}}-1 }  \nonumber\\
&-&\,\delta (z^{+}) \, \frac{1}{(2 \pi) ^{2}L} \,\int \, d^{2} k_{\perp} \; \, \frac{ e^{- i k_{\perp} z^{\perp} }}{  k^{2}_{\perp}- i \epsilon} \;.  
\end{eqnarray}
$D_{lc}$ contains a singular contribution arising from the zero-mode (n=0 in Eq.(\ref{d3})). Up to this zero-mode contribution the light-cone two-point function agrees with the infinite momentum frame limit of the two-point function $D_{-}$ (Eq.(\ref{exp})) provided  the light-cone extension $L$ is identified with the Lorentz-contracted extension $\sqrt{1+\beta } \, L$ in the infinite momentum frame which  becomes small on the scale of the regulator.
In the longitudinal Casimir effect,  the compact direction is specified by the lightlike vector given in light-cone coordinates by 
$$\ell = (0,L,0,0) \, ,\quad \ell^{2} = 0.$$ 
Therefore the two-point function  
$$D_{lc}= D_{lc}(z^2, z\ell ),$$
cannot contain a finite and regulator independent term. A meaningful definition of the energy density for the compact direction  coinciding with the $x^{-}$ direction is not possible. However the compact direction can be arbitrary close to the $x^{-}$ direction. This is easily seen when 
 imposing   boundary conditions  
\begin{equation}
  \label{bcp}
  \varphi(x^+,x^-+L,x^1+sL,x^2)= \varphi(x^+,x^-,x^1,x^2) \;,
\end{equation}
by which the $x^{-}$ direction can be approached ($s \rightarrow 0$) from orientations characterized by spacelike vectors
$$\ell = (0,L,sL,0)\, ,\quad \ell^{2} = -s^2 L^2 . $$
By a Lorentz-transformation consisting of a rotation around the $x^2$ - axis by an angle $\alpha=\arcsin (1+2s^2)^{-1/2}$ followed by a boost along the $x^1$ direction with velocity $\beta =\sin \alpha$ transforms this boundary condition into  a transverse boundary condition
\begin{equation}
  \label{bcpp}
  \varphi(x^+,x^-,x^1+sL,x^2)= \varphi(x^+,x^-,x^1,x^2) .
\end{equation}
The expansion of $D(z)$ in Eq.(\ref{expa}) is controlled by  $z/(sL)$ which for given s can be made arbitrarily small in the $z \rightarrow 0$ limit. In this limit,   
the light-cone energy density is obtained from Eq.(\ref{energy2})
$$ \langle J_{+-}\rangle = - \,\frac{\pi^2}{90 (sL)^4}\, .$$
Thus a physically meaningful  Casimir energy emerges for arbitrary choices of the compact spacelike direction excluding  a  region around the $x^{-}$ direction with opening angle determined by the ratio of the components of the regulator $z$ over the proper length $sL$. 
\vskip 0.2cm
\,  The failure   
 of the regularization to produce sensible results for a compact $x^{-}$ direction is  due to the peculiar infrared properties of the spectrum in light-cone quantization.  It is the divergence of the one particle energies at small light-cone momenta which is regularized by $z^{+}$ and which makes the final result dependent on this regulator. On the other hand, after subtraction of the ultra\-violet contribution, the Casimir effect is a long wavelength phenomenon as the difference  between  the characteristic $L^{-4}$ dependence  for massless and the exponential suppression $\exp{(-mL)}$ for massive particles demonstrates. Furthermore the attraction, i.e. the decrease in the energy density with the decrease in the separation of the walls is a result of  a delicate interplay between the repulsion due to the increased zero-point energies for $n\neq 0$  and the increase in the relative weight of the states with small or vanishing $n$. As a consequence of these compe\-ting effects a change from attraction to repulsion occurs when chan\-ging continously from periodic to antiperiodic boundary conditions. The lightlike nature of the compact direction in combination with this sensitivity of the observables to the infinite and long wavelength properties makes the energy-momentum tensor in the longitudinal Casimir effect ill-defined. Similar difficulties are most likely to be encountered in the attempt to calculate other vacuum properties such as condensates which also are dominated by long wavelength properties.  Like in the Casimir effect these problems can be avoided provided  compact directions, if at all present, are chosen to be spacelike. The choice of  spacelike compact directions raises however  new technical issues.  Compactification of the $x^{-}$ direction is often the basis for significant simplifications in the actual calculations. In gauge theories, for instance,  the choice of light-cone gauge $A_{-}$ seems to be a prerequisite for making use of the simplifications offered by light-cone quantization. On the other hand, implementation of axial gauges is free of infrared problems only if the associated direction - here the $x^{-}$ direction - is compac\-tified. One therefore again may have to resort to approach the lightlike from spacelike compact directions which in turn would require an accompanying change in gauge. These problems have to be investigated  if va\-cuum properties are to be determined within approaches to light-cone quantizations which make use of a compact $x^{-}$ direction such as  DLCQ \cite{pauli1,pauli2,pauli3,pauli4} or the transverse lattice approach \cite{bardeen,burkardt,dalley,sande}.   

\vskip 0.2cm
\, Extensions to other forms of boundary conditions are possible. Quasi-periodic boundary conditions are important because of their application to finite temperature field theory. If one requires
\begin{equation}
  \quad \varphi (x+\ell) = e^{i\chi}\varphi (x) \,\,\,,0\leq \chi < 2\pi \,\,,
\end{equation}
the analysis can be carried through as before. The expression (\ref{energy2}) for the $L$-dependent part of the energy momentum tensor gets modified by
\begin{align}
\langle J_{\mu\nu}\rangle \rightarrow \langle J_{\mu\nu}\rangle\cdot \left(1-\frac{15\chi^2}{2\pi^2}\left(1-\frac{\chi}{2\pi}\right)^2\right) \, .
\end{align}
In the light-cone result (\ref{lcl2}) for the longitudinal Casimir effect, the relevant term in $\langle {\cal H}_{lc} \rangle$ becomes
\begin{align}
\frac{1}{24 \tilde{L}^4} \rightarrow \frac{1}{24 \tilde{L}^4} \cdot \left(1-\frac{3\chi}{\pi}\left(1-\frac{\chi}{2\pi}\right)\right) \, .
\end{align}
Therefore the difference between our results for periodic and quasi-periodic boundary conditions consists only of some $\chi$-dependent factors which do not influence our conclusions with respect to light-cone physics. 
Extension to standing wave boundary conditions like
\begin{align}
\quad \varphi(\mbox{\boldmath$x$}_{\perp},x^3=0)=0, \, \varphi(\mbox{\boldmath$x$}_{\perp},x^3=L)=0\,,
\end{align}
introduce new elements in our discussion. First we remark that with the violation of translational invariance by the Dirichlet boundary conditions, the energy momentum tensor becomes $x^3$-dependent. Following the same procedure as above we find e.g. for the energy density as seen from a moving observer
\begin{align}
\label{stand}
\quad \langle {\cal H}\rangle  &= -\frac{\pi^2}{90(2L)^4} [1+4\beta^2\gamma^2] \nonumber\\
&- \frac{\pi^2}{24 L^4} \frac{2+\cos\left(\frac{2\pi}{L}\tilde{x}^3\right)}{\left(1-\cos\left(\frac{2\pi}{L}\tilde{x}^3\right)\right)^2} \nonumber\\
&\;\;\cdot[2+3\beta^2\gamma^2] \Theta(\tilde{x}^3)\Theta\left(L-\tilde{x}^3\right)
\end{align}
with $ \tilde{x}^3=\gamma(x^3-\beta x^0)$. 
After integration over $x^3$, the second term in (\ref{stand}) becomes $L$-independent and therefore does not contribute to the force between the walls. In comparison with periodic boundary conditions the well-known $2^4$ suppression of the force generated by standing waves is obtained. Once more the transverse Casimir effect on the light-cone yields the same result. Although the Casimir effect is well defined in the infinite momentum frame limit ($\beta^2 \rightarrow 1$ in(\ref{stand})), no physically meaningful description for the longitudinal Casimir effect in light-cone quantization exists. Equal light-cone time standing wave conditions are not compatible with the light-cone equations of motion which are first order in $x^-$.


\section{Thermodynamical Observables in Light-Cone Quantization}
The correct description of the Casimir effect for spacelike compact directions offers the possibility to calculate thermodynamic quantities in light-cone quantization. The straightforward generalization of the standard procedure to define a partition function in light-cone quantization by compactifying  the  light-cone time is faced with  the difficulties encountered in the definition of the longitudinal  Casimir effect.  The light-cone time direction is lightlike  and not as in ordinary coordinates timelike. Compactification along the $x^{+}$ direction at inverse temperature $\beta$ defines the 4-vector with light-cone coordinates 
$$\ell=(\beta,0,0,0),\quad \ell^{2}=0$$
implying 
$$D(z)=D(z^2,z^{-}\beta).$$
A finite regulator independent thermodynamic quantity cannot be extracted from $D(z)$. 
We however may use the general equivalence between relativistic field theories at finite temperature and finite extension \cite{lt98,toms}. 
By rotational invariance in the Euclidean, the value of the
partition function of a system with finite extension $L$
in 1 direction and $\beta$  in 0 direction is invariant under the
exchange
of  these two extensions,
\begin{equation}
Z\left(\beta,L\right)=Z\left(L,\beta\right) \ ,
\label{FE1}
\end{equation}
provided bosonic (fermionic) fields satisfy periodic (antiperiodic)
boundary conditions in both time
and 3 coordinate.  Thus relativistic covariance connects the
thermodynamic properties of a canonical ensemble with 
vacuum properties of  the same physical system but at finite extension. As a consequence 
 energy density and pressure are related
by
\begin{equation}
\epsilon\left(\beta,L\right)=-p\left(L,\beta\right) \ .
\label{FE2}
\end{equation}
Thus  the energy-momentum tensor at finite temperature is trivially computed once the
 energy-momentum tensor at finite extension is known. As we have shown these elements can be evaluated on the light-cone provided a spacelike compact direction is chosen. The general relation between systems at finite temperature and finite extension  not only connects the Casimir effect of a non-interacting massless field with blackbody radiation it also implies the possibility for phase transitions to occur with the variation in the extension $L$ of e.g. the compact $x^{1}$ direction. Specifically in QCD,  when $L$ decreases beyond $\sim 1.3 $fm the phase transition to the quark-gluon plasma has to take place with a corresponding sudden change in energy density and pressure. Thus light-cone quantization provides an appropriate framework in which phase transitions at finite temperature can be described. Nevertheless  a detailed understanding of  how  such phase transitions arise in the trivial light-cone vacuum remains to be achieved.
\section{Summary}
In this work we have discussed the Casimir effect of a non-interacting scalar field in light-cone quantization. We have established that the Casimir effect can be reliably computed provided the separation of the walls enclosing the system is  spacelike. This successful evaluation opens the possibility to calculate thermodynamic quantities  and in particular to address the issue of phase transitions in finite temperature field theory in the framework of light-cone quantization. If the walls are separated along the $x^{-}$ direction no regulator independent expression for the observable pressure can be obtained. Based on the covariance of the theory  this failure in defining a proper energy momentum tensor has been shown to result from the  lightlike separation along the $x^{-}$ axis. Therefore the same problems will  also occur for interacting fields. Compact  light-like directions might be approached in some limiting procedure from compact space-like directions. We have studied this possibility by the transition to the infinite momentum frame and by a rotation of a spacelike orientation into the lightlike direction. In both cases the final result has been demonstrated to depend on the order in which the approach to the light-like direction and the limit in the regularization of the quantum fluctuations are performed. We have not discussed here the possibility to resolve this problem by approaching the light-cone metric from a metric with 3 spacelike coordinates. This approach to light-cone quantization has been proposed in the field-theoretic context in \cite{FR,ltly91} and in the context of M-theory in \cite{POL99,bilal1,bilal2}. It has been applied  to a detailed analysis of vacuum properties of 2-dimensional gauge theories \cite{ltly91}. This limiting procedure  also yields for the longitudinal Casimir effect the correct result \cite{Stei01}. Here we have not followed this path since in general   most of  the simplifying features of light-cone quantization are thereby lost.  Beyond two dimensions this method is not mandatory. It is not the choice of the metric  but rather the choice of a compact $x^{-}$ direction which is the origin of the difficulties in defining the  Casimir effect.  
\vskip 0.5cm
\noindent {\bf Acknowledgments} 
We thank M. Burkardt and M. Thies for illuminating discussions.
\vskip 0.5cm 
\
          
 \end{document}